\documentclass[apj]{emulateapj}

\begin{document}


\title{Ambipolar Electric Field, Photoelectrons, and their Role in Atmospheric Escape From Hot-jupiters}


\author{O. Cohen\altaffilmark{1}, A. Glocer\altaffilmark{2}}

\altaffiltext{1}{Harvard-Smithsonian Center for Astrophysics, 60 Garden St. Cambridge, MA 02138, USA.}
\altaffiltext{2}{NASA/GSFC, Code 673 Greenbelt, MD 20771, USA.}

\begin{abstract}

Atmospheric mass-loss from Hot-jupiters can be large due to the close proximity of these planets to their host star and the strong radiation the planetary atmosphere receives. On Earth, a major contribution to the acceleration of atmospheric ions comes from the vertical separation of ions and electrons, and the generation of the ambipolar electric field. This process, known as the "polar wind", is responsible for the transport of ionospheric constituents to the Earth's magnetosphere, where they are well observed. The polar wind can also be enhanced by a relatively small fraction of super-thermal electrons (photoelectrons) generated by photoionization. We formulate a simplified calculation of the effect of the ambipolar electric field and the photoelectrons on the ion scale-height in a generalized manner. We find that the ion scale-height can be increased by a factor of 2-15 due to the polar wind effects. We also estimate a lower limit of an order of magnitude increase of the ion density and the atmospheric mass-loss rate when polar wind effects are included.

\end{abstract}


%%%%%%%%%%%%%%%%%%%%%%%%%%%%%%%%%%%%%%%%%%%%%%%%%%%%%%%%%%%%%%%%%%%%%%%%%%%%%%
% Section - Introduction
%%%%%%%%%%%%%%%%%%%%%%%%%%%%%%%%%%%%%%%%%%%%%%%%%%%%%%%%%%%%%%%%%%%%%%%%%%%%%%

\section{INTRODUCTION}
\label{sec:Intro}

In the past two decades and in particular, following the {\it Kepler} mission, hundreds of exoplanets have been detected \citep[e.g.,][]{exoplanet95,exoplanets03}. Many of these planets are gas giants observed at an extremely close orbit of less than 0.1~AU from their host star (an orbital period of less than 10 days), and are classified under the term "Hot-Jupiters" (HJ). The unexpected close-in orbit of HJ has stimulated many science investigations regarding their formation, evolution, and tidal interaction \citep[e.g.,][and references therein]{Papaloizou07}, their magnetic interaction with the host star \citep[e.g.,][and references therein]{Cohen10}, and the structure and dynamics of their atmospheres \citep[e.g.,][and references therein]{Showman08}.

In such a close orbit (especially if the star and the planet are tidally-locked), HJ are expected to receive extremely large amounts of stellar X-ray and EUV radiation \citep{Penz08,Cecchi-Pestellini09}. It has been argued that this high EUV radiation can lead to a strong photo-evaporation of the planetary atmosphere and high mass loss rates \citep{Lammer03,Baraffe04, Baraffe06}, leading to a less massive planets. However, this could not be supported by the observed mass distribution \citep{Hubbard07}. Observations of $Ly\alpha$ emission from the HD209458 system have suggested that the planet occupies an inflated hydrogen corona with outflow velocities of $50-100\;km\;s^{-1}$ and a mass-loss rate of about $10^{10}\;g\;s^{-1}$. However, there is a debate on whether the observations are effected by the host star or whether the observed features are of planetary origin \citep{Vidal-Madjar03,Ben-Jaffel07,Vidal-Madjar08}. A more recent observation of the system, as well as of the HD189733 system reveled a smaller mass-loss rate of about $10^{8}\;g\;s^{-1}$ \citep{LecavelierDesEtangs10,Linsky10}.

On the theoretical side, several models for atmospheric escape from HJ have been developed in recent years. A detailed models for the chemistry, photoionization, and aeronomy of HJ were developed by \cite{Yelle04} and by \cite{Garcia-Munoz07}. \cite{Tian05} and \cite{Murray-clay09} performed hydrodynamic calculations of thermally driven atmospheric escape, and \cite{Stone09}, \cite{Trammell11}, and \cite{Adams11} included the planetary magnetic field geometry, which confines the escaping gas to regions of open field lines. The models above predict mass-loss rates not higher than $10^{10}\;g\;s^{-1}$. Some of the models also included the incoming stellar wind and found that the planetary outflow is ought to be suppressed by the wind. Non of the models predicts a sufficiently high mass-loss rate so that the planet can be evaporated in a relatively short time-scale.

In the Earth's upper atmosphere (as well as in other planets), there is a well observed physical process which plays an important role in the acceleration of ions. The polar wind \citep{BanksHolzer68} is the outflow of planetary ions along open field lines. The main driver for this process is the ambibolar electric field, which is proportional to the electron pressure gradient. Since electrons are more mobile than ions, a charge separation is created along the magnetic field direction, leading to an electric potential that acts on the ions to retain charge neutrality. The end result is an acceleration of the ions by this electric field so that the ions are dragged by the electrons. 

The electric field applies a force propotional to the negative gradient of the electron pressure. Using some simplifications, the resulting force is approximately equivalent to half of the gravitational force on the major ion species and directed oppositely. Since $O^{+}$ is the major ion species in Earth's upper ionosphere, the result is a supersonic flow of H$^{+}$ and an increase in the O$^{+}$ scale height. In addition, photoelectrons, which are highly energized electrons due to photoionization (the tail of the distribution function), can significantly increase the electron temperature, leading to an enhancement of the ions acceleration \citep{Lemaire72}. In the Earth's upper atmosphere, the velocity of $O^+$ is lower than the escape velocity. Nevertheless, $O^+$ is observed to serve as a significant plasma source in the magnetosphere \citep{Lennartsson81}. \cite{Tam95} and \cite{Tam98} have demonstrated by numerical simulation that photoelectrons indeed, can accelerate $O^+$ and $H^+$, while they obtained an unrealistic electron temperature of $40,000\;K$. An additional simulation by \cite{khazanov97} resulted in a more realistic electron temperature of $16,000\;K$. Recent numerical simulations by \cite{Glocer12} also included the effects of photoelectrons to look at the global outflow solution and compared with in-situ observations. Their simulations showed that the polar wind mechanism is responsible to the transport of ionospheric $H^+$ and  $O^+$, and that only a small fraction of photoelectrons can significantly contribute to the ion acceleration.

In this letter, we investigate how the ambipolar electric field and the fraction of photoelectrons reduce the gravitational potential, and therefore, increase the ion scale-height and the ion density at the top of the atmosphere of HJ. We also calculate how the mass-loss rate for $H^+$ is modified by this effect. Due to the high EUV radiation, the fraction of photoelectron in the atmospheres of HJ is expected to be higher than in the Earth's case, leading to a much greater increase of electron temperature. 

In Section~\ref{sec:AEF}, we calculate the change of the effective gravity and the ion scale-height due to the ambipolar electric field and photoelectrons. We present and discuss the results in Section~\ref{sec:Results}, and draw our conclusion in Section~\ref{sec:Conclusions}.

%%%%%%%%%%%%%%%%%%%%%%%%%%%%%%%%%%%%%%%%%%%%%%%%%%%%%%%%%%%%%%%%%%%%%%%%%%%%%%
% Section - Simulation
%%%%%%%%%%%%%%%%%%%%%%%%%%%%%%%%%%%%%%%%%%%%%%%%%%%%%%%%%%%%%%%%%%%%%%%%%%%%%%
\section{MODIFICATION OF THE ION SCALE-HEIGHT BY THE AMBIPOLAR ELECTRIC FIELD }
\label{sec:AEF}

In the derivation bellow, we follow the standard model for the polar wind, but we include the effect of the photoelectrons on the solution. For a planetary atmosphere consiststing of electrons, photoelectrons, and ions, charge neutrality requires that:
\begin{equation}
n_{e0}+n_{\alpha 0}=n_{i0},
\label{neutrality}
\end{equation}
where $n_{e0}$, $n_{\alpha 0}$, and $n_{i0}$ are the electron, photoelectron, and ion number densities at some reference altitude, $r_0$. From Eq.~\ref{neutrality}, we can define the fraction of the photoelectrons, $\beta$, as $n_{\alpha 0}=\beta n_{i0}$, and the fraction of electrons as $n_{e 0}=(1-\beta)n_{i0}$. 

Our goal here is to calculate how the effective gravity at the surface is modified when taking into account the photoelectrons and the ambipolar electric field, and investigate how this modified gravity affects the ion scale-height $H_i$. We will compare $H_i$ with the unchanged scale-height $H_0$ which contains the {\it surface} gravity $g$ but not the ambipolar electric field. 

We begin by assuming a hydrostatic ion density profile: 
\begin{equation}
n_i(z)=n_{i0}e^{-(z-z_0)/H_i},
\label{ionprofile}
\end{equation}
with the ion scale-height, $H_i=\frac{kT_i}{m_ig_{eff}}$, where $k$ is the Boltzmann constant, $T_i$ is the ion temperature, $m_i$ is the ion mass, and $g_{eff}$ is the effective gravity. Without the effects we study here, $g_{eff}=g$. Conservation of the photoelectron mass along a magnetic flux-tube requires that:
\begin{equation}
n_{\alpha 0}u_{\alpha 0}A_0=n_{\alpha}u_{\alpha}A,
\label{photoelectroncont1}
\end{equation}
with $u_{\alpha 0}$ and $u_{\alpha}$ being the photoelectrons velocities, and  $A_0$, and $A$ being the magnetic flux tube cross-sections at the reference altitude and at some altitude, respectively. This equation implicitly neglects any scattering of the photoelectrons. In a magnetic dipole geometry, the magnetic flux conservation requires that $A_0B_0=AB$, with $B=C/r^3$ being the dipole field magnitude as a function of radius ($C$ is a constant), and $B_0=C/r^3_0$ is the field magnitude at the reference altitude, $r_0$. Therefore, $A_0/A=r^3_0/r^3$, and we have:
\begin{equation}
n_{\alpha}=n_{\alpha 0}\left( \frac{r_0}{r} \right)^3,
\label{photoelectroncont2}
\end{equation}
assuming $u_{\alpha 0}=u_\alpha$ as a lower limit. Using Eq.~\ref{ionprofile} and \ref{photoelectroncont2}, the electron density at altitude $z$, can now be obtained, assuming $z_0=0$, $r_0=R_p$, and $r=R_p+z$:
\begin{equation}
n_e(z)=n_i(z)-n_\alpha(z)=n_{i0}\left( e^{-z/H_i}-  \frac{\beta R^3_p}{(R_p+z)^3} \right).
\label{electronden}
\end{equation}

The effective gravity is modified by the ambipolar electric field as $g_{eff}=g-\frac{eE_\parallel}{m_i}$, with the ambipolar electric field ({\it positive for ions}) defined as \citep{SchunkNagy04}:
\begin{eqnarray}
E_\parallel=\frac{1}{e n_e}\frac{\partial p_e}{\partial z}=\frac{kT_e}{e n_e}\frac{\partial n_e}{\partial z}=\nonumber\\
\frac{kT_e n_{i0}}{en_e}\left[  -\frac{1}{H_i}e^{-z/H_i} +\frac{3\beta R^3_P}{(R_p+z)^4} \right].
\label{Epar}
\end{eqnarray}
Here $T_e$ is the electron temperature, and $e$ is the electric charge. At the planetary surface, $z=0$ and so we obtain:
\begin{equation}
E_\parallel(z=0)=-\frac{kT_e }{e(1-\beta)}\left(  -\frac{1}{H_i}+\frac{3\beta}{R_p} \right),
\end{equation}
which yields:
\begin{equation}
g_{eff}= g+\frac{kT_e }{m_i(1-\beta)}\left(  -\frac{m_ig_{eff}}{kT_i}+\frac{3\beta}{R_p} \right) 
\end{equation}
or:
\begin{equation}
g_{eff} =\left [g+\frac{3\beta kT_e}{m_iR_p(1-\beta )} \right] \left[ \frac{(1-\beta)T_i}{T_e+(1-\beta)T_i} \right]
\label{effectiveG}
\end{equation}
In Eq.~\ref{effectiveG}, $g$ is modified by the ion and electron temperatures, and by the fraction of photoelectrons. For the case of $\beta=0$ and $T_i=T_e$, the well-known reduction of the effective gravity of the ions by half is obtained \citep{Gombosi04}. 

As shown by previous models \citep{Tam95,khazanov97,Tam98,Glocer12}, the electron temperature is highly affected by even a very small fraction of photoelectrons. In our model here, we assume that $T_i=1000K$. Despite of the higher ion temperature expected in HJ, the effect studied here is driven by the difference between $T_e$ and $T_i$, so that it should scale with the increase in $T_i$. We scale the electron temperature with the percentage of photoelectrons and $T_i$ using two different models. One is based on the electron temperature distribution at the top of the Earth's atmosphere from \cite{khazanov97}:
\begin{equation}
T_e(\beta)=T_i*2^{6+\log{\beta}},
\label{Te1}
\end{equation}
with $1000<T_e<16000K$ for $10^{-6}(10^{-4}\%)<\beta<10^{-2}(1\%)$, and a more modest function with $1000<T_e<10000K$:
\begin{equation}
T_e(\beta)=T_i*1.8^{6+\log{\beta}}.
\label{Te2}
\end{equation}
With the above models for $T_e$, the modified gravity and scale-height can be calculated as a function of the fraction of photoelectrons.

%%%%%%%%%%%%%%%%%%%%%%%%%%%%%%%%%%%%%%%%%%%%%%%%%%%%%%%%%%%%%%%%%%%%%%%%%%%%%%
% Section - Results
%%%%%%%%%%%%%%%%%%%%%%%%%%%%%%%%%%%%%%%%%%%%%%%%%%%%%%%%%%%%%%%%%%%%%%%%%%%%%%

\section{RESULTS \& DISCUSSION}
\label{sec:Results}

\subsection{Results}

\begin{figure*}[h!]
\centering
\includegraphics[width=5.5in]{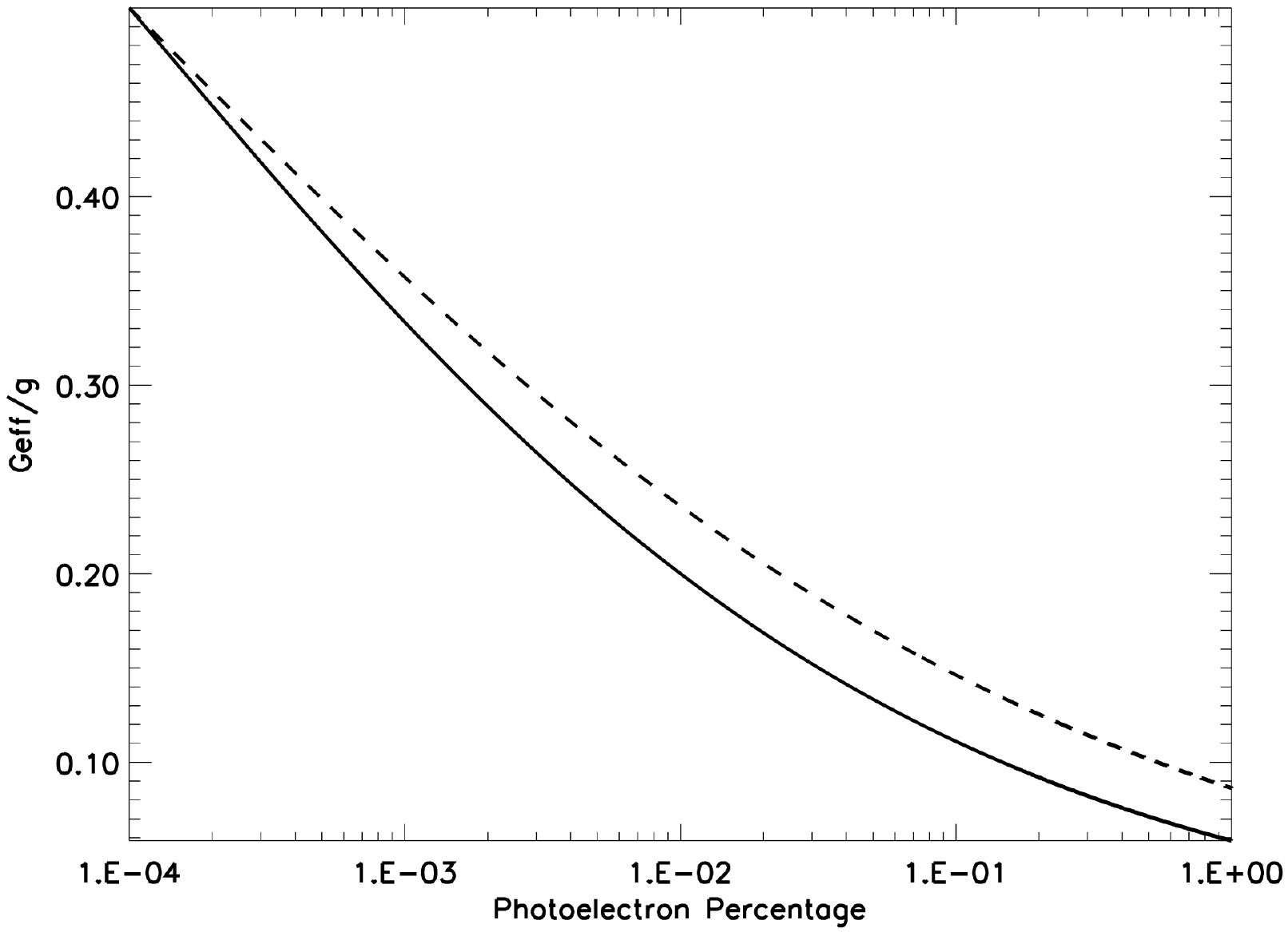}
\caption{Effective gravity as a function of the photoelectron percentage ($0.0001-1\%$) for $T_e$ model 1 (solid line) and model 2 (dashed line).}
\label{fig:f1}
\end{figure*}

\begin{figure*}[h!]
\centering
\includegraphics[width=5.5in]{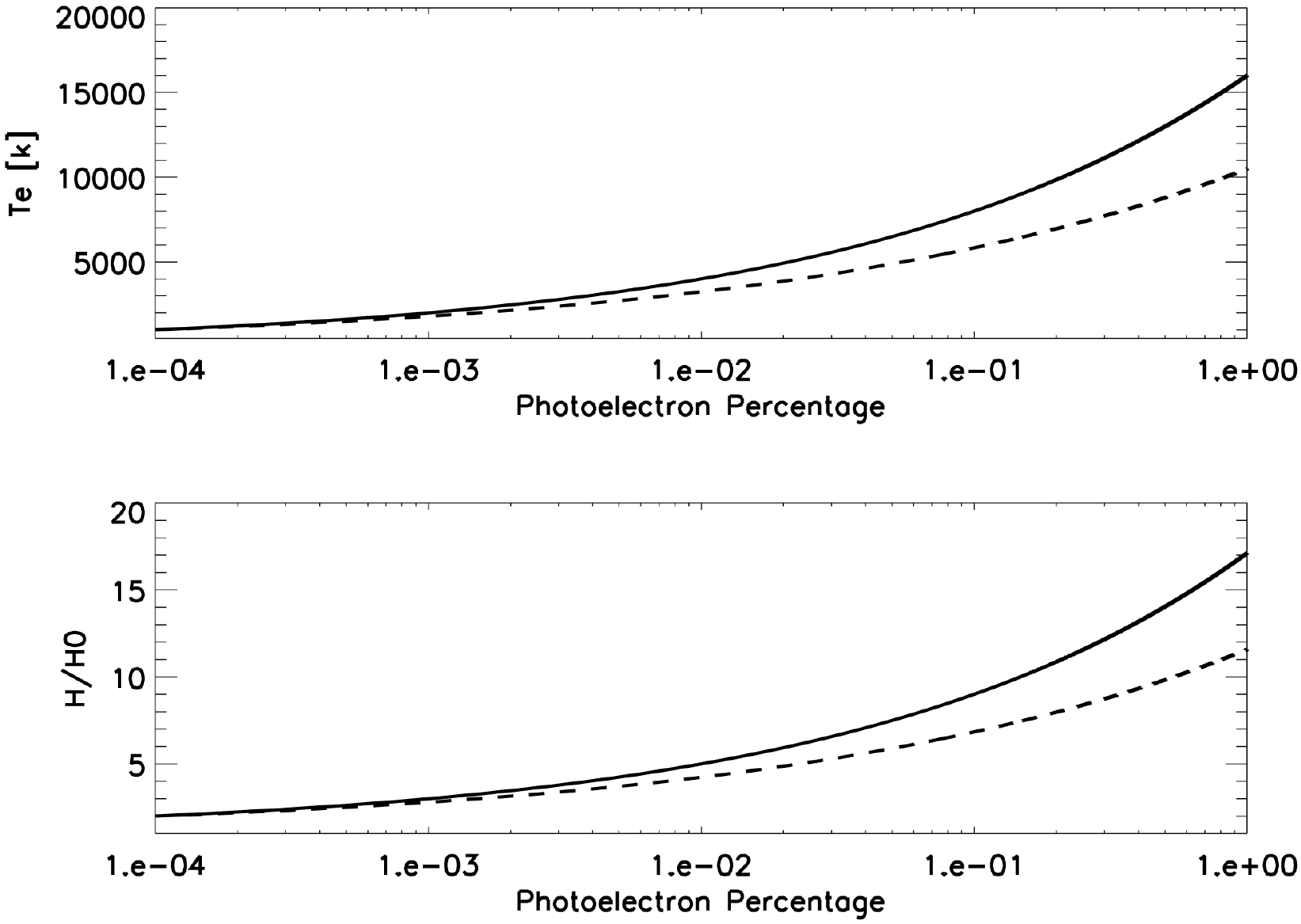}
\caption{Electron temperature (top) and the ratio of modified to unmodified ion scale-height (bottom) as a function of the photoelectron percentage ($0.0001-1\%$) for $T_e$ model 1 (solid line) and model 2 (dashed line).}
\label{fig:f2}
\end{figure*}

Figure~\ref{fig:f1} shows the effective gravity as a function of the photoelectron percentage for the two models for $T_e$. For $\beta=0$, we obtain $g_{eff}/g=0.5$. In Figure~\ref{fig:f2}, we show the electron temperature and the ratio of modified to non-modified scale-height as a function of the fraction of photoelectrons, assuming $m_i=m_p$, the proton mass. Here, we show the solution only for photoelectron percentages of $0.0001-1$. One can see that if the fraction of photoelectrons is even less than 1\%, the scale-height increases by a factor of 2-15. 

\begin{figure*}[h!]
\centering
\includegraphics[width=5.5in]{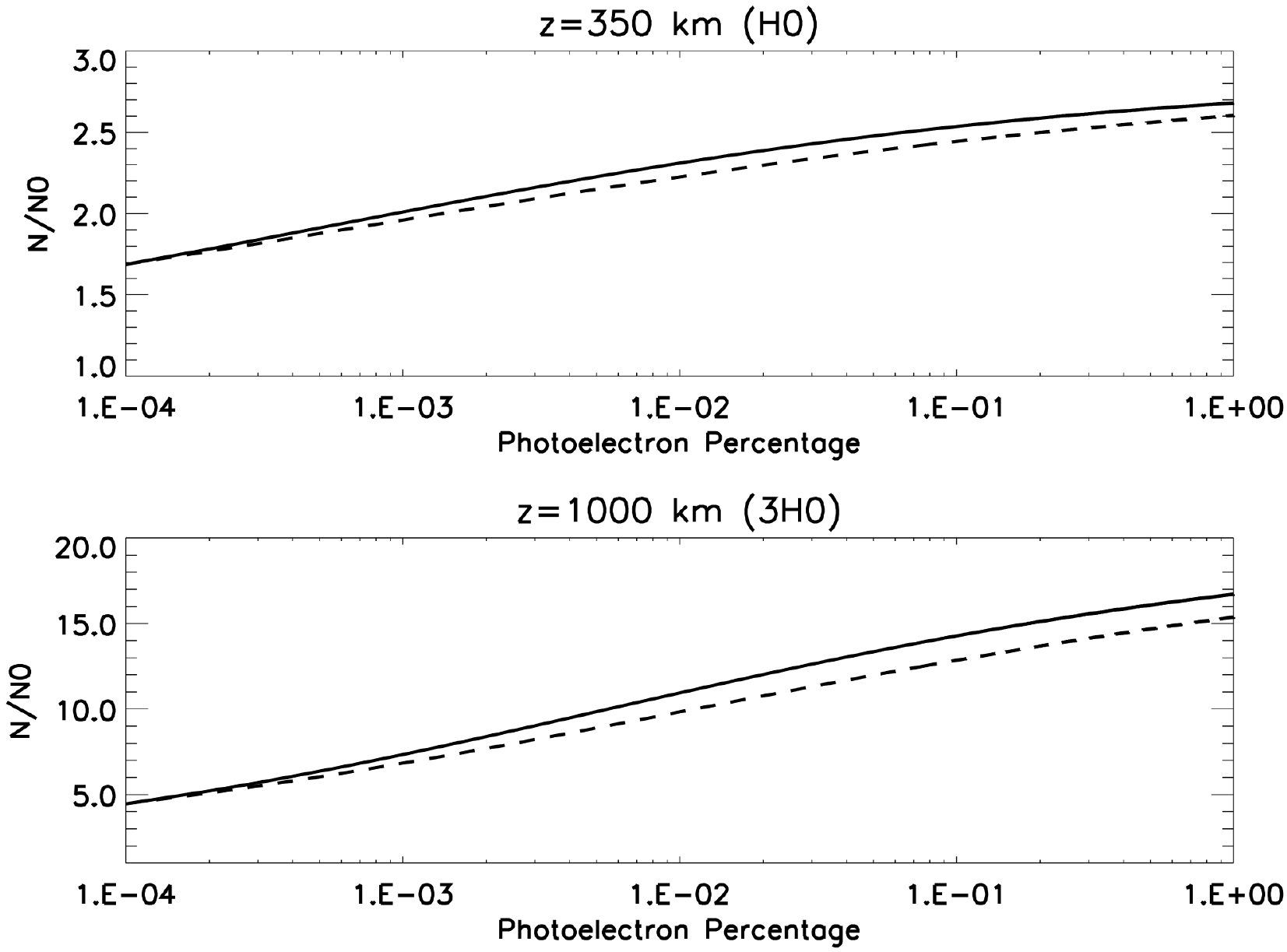}
\caption{Ratio of non-modified to modified hydrostatic ion densities for $T_e$ model 1 (solid line) and model 2 (dashed line) at $z=350\;km$ (top) and at $z=1000\;km$ (bottom) as a function of the photoelectron percentage ($0.0001-1\%$).}
\label{fig:f3}
\end{figure*}

A realistic ion density profile cannot be obtained using the simplified calculation we present here. In particular, we cannot calculate the density profile of the $H^+$ ions, since they are expected to attain supersonic speeds. Therefore, it is hard to estimate the increase in density at the top of the atmosphere and the corresponding increase in mass-loss rate. Nevertheless,  we can use a hydrostatic profile to estimate the ion density change at lower altitudes. In Figure~\ref{fig:f3}, we show the ratio of the hydrostatic density profiles using the modified and unmodified scale-heights, respectively, as a function of the fraction of photoelectrons for altitudes of $350\;km$ ($\sim 1H_0$) and $1000\;km$ ($\sim 3 H_0$). The density is increased by a factor of 2-3 at $350\;km$ and by a factor of 5-15 at $1000\;km$. At higher latitudes, the hydrostatic solution is probably not valid anymore and the ratio in Figure~\ref{fig:f3} will become too large, since the density profile for the unmodified scale-height goes to zero faster than the one with the modified scale-height. 

\subsection{Discussion}

In HJ, the extremely strong radiation is expected to increase the fraction of photoelectrons. Therefore, the electron temperature should be higher than the ion temperature, despite of the strong heating at the day side, so that the mechanism proposed here should still be significant. The effect should be limited at the night side due to the lower ionization rate, and it is not clear yet how effective the atmospheric day-night circulation is at higher latitudes (where the day-night temperature difference is smaller than that at the equator), and at high altitudes (where the ion acceleration occurs). 

For a magnetized HJ, the mass-loss is expected to take place along the magnetic field lines which are open to the stellar wind \citep[as demonstrated by][]{Stone09,Trammell11,Adams11}, and that is exactly where the polar wind process takes place. It has been previously shown that the classical polar wind mechanism together with the addtion of photoelectrons and wave-particle interactions is responsible for the transport of $H^{+}$ and $O^{+}$ out of the Earth's atmosphere. By lowering the potential barrier, these processes effectively lower the the escape velocity. These processes have also been speculated to be important at Jupiter and Saturn \citep{Glocer07, Nagy86}; the major ions in the upper atmosphere at these planet are $H_{3}^{+}$ and $H^{+}$. In HJ, the relative ion abundances are not known, but modeling by \citet{Garcia-Munoz07} shows that $H^{+}$, $H_{3}^{+}$, $He^{+}$, $C^{+}$, and various ionized hydrocarbons are possibly present. The polar wind process should apply to each of these planets. Indeed the derivations presented here reflects the basic textbook deriviation of the classical polar wind \citep{Gombosi04} to which we have added the effect of photoelectons. No other planet specific parameters are required. Even neglecting the effect of photoelectrons, the polar wind process by itself could significantly increase the ion scale height. 

The relative composition affects the polar wind process by changing the parallel electric field. This is because the parallel electric field was found to increase with mass. In the case of no photoelectrons, if $H_{3}^{+}$ was the major ion species (such as at Jupiter or Saturn) than the parallel electric field would excert an upward force approximately equal to one half the gravitation force acting on $H_{3}^{+}$. In this case the scale height of $H_{3}^{+}$ would increase. Lighter consituents such as $H^{+}$ would actually have a net upward force resulting in eventual supersonic flow. Including photoelectrons, increases the electric field and intensifies the effect of the polar wind, possibly resulting in a net upward force on heavier species. If $H^{+}$ was the major ion, the parallel electric field would be reduced, but the effect would still be quite significant. 

The simplified model presented here is insufficient to predict the detailed change in the ion density profile, but it can predict how the scale height changes. We show that this change can reach about a factor of 10 at lower altitudes. Therefore, it should also increase the mass-loss rate by the same amount assuming the same surface area, and without changing the ion velocity at the top of the atmosphere. The polar wind is expected to further accelerate the ions such that, the ion speed should increase as well, so the factor of 10 increase is a lower limit. 

In order to perform a more detailed calculation of the effect of the polar wind on the mass-loss rate of HJ, a more detailed model is needed, such as the polar wind model by \cite{Glocer07}, \cite{Glocer09}, and \cite{Glocer12}, which is similar to that of \cite{Garcia-Munoz07}, but includes the effect of the ambipolar electric field and photoelectrons. The derivation and discussion contained in this paper, however, demonstrates that the polar wind process plays an important role in the mass-loss rate of HJ and should be accounted for in models.  

%%%%%%%%%%%%%%%%%%%%%%%%%%%%%%%%%%%%%%%%%%%%%%%%%%%%%%%%%%%%%%%%%%%%%%%%%%%%%%
% Section - Conclusions
%%%%%%%%%%%%%%%%%%%%%%%%%%%%%%%%%%%%%%%%%%%%%%%%%%%%%%%%%%%%%%%%%%%%%%%%%%%%%%
\section{Conclusions}
\label{sec:Conclusions}

In this paper, we perform a simplified calculation of the effect of the ambipolar electric field and atmospheric photoelectrons on the planetary ion scale-height. We show that this effect can reduce the effective gravity and therefore, enhance the ion acceleration in the region of the planetary atmosphere, where magnetic field lines are open. We find that a small fraction of photoelectrons (less than 1\% of the total electrons) can increase the ion scale-height by a factor of 2-15. We calculate the hydrostatic density profiles using the modified scale-heights and find that the planetary mass-loss rate should increase by an order of magnitude at a minimum, even neglecting any increase in the ion velocity due to this the process. Since the ion acceleration should be enhanced by the process, we expect the increase in mass-loss rate to be even greater. A more comprehensive calculation, however, requires a more detailed modeling effort.  

%%%%%%%%%%%%%%%%%%%%%%%%%%%%%%%%%%%%%%%%%%%%%%%%%%%%%%%%%%%%%%%%%%%%%%%%%%%%%%
% Acknowledgments
%%%%%%%%%%%%%%%%%%%%%%%%%%%%%%%%%%%%%%%%%%%%%%%%%%%%%%%%%%%%%%%%%%%%%%%%%%%%%%

\acknowledgments
We thank an unknown referee for her/his review report, and Jeremy Drake for his help in preparing this manuscript. OC is supported by SI Grand Challenges grant number 40510254HH0022. 

%%%%%%%%%%%%%%%%%%%%%%%%%%%%%%%%%%%%%%%%%%%%%%%%%%%%%%%%%%%%%%%%%%%%%%%%%%%%%%
% Bibliography
%%%%%%%%%%%%%%%%%%%%%%%%%%%%%%%%%%%%%%%%%%%%%%%%%%%%%%%%%%%%%%%%%%%%%%%%%%%%%%

%\bibliographystyle{apj}
%\bibliography{AE.bib}

\begin{thebibliography}{35}
\expandafter\ifx\csname natexlab\endcsname\relax\def\natexlab#1{#1}\fi

\bibitem[{{Adams}(2011)}]{Adams11}
{Adams}, F.~C. 2011, \apj, 730, 27

\bibitem[{{Banks} \& {Holzer}(1968)}]{BanksHolzer68}
{Banks}, P.~M., \& {Holzer}, T.~E. 1968, \jgr, 73, 6846

\bibitem[{{Baraffe} {et~al.}(2006){Baraffe}, {Alibert}, {Chabrier}, \&
  {Benz}}]{Baraffe06}
{Baraffe}, I., {Alibert}, Y., {Chabrier}, G., \& {Benz}, W. 2006, \aap, 450,
  1221

\bibitem[{{Baraffe} {et~al.}(2004){Baraffe}, {Selsis}, {Chabrier}, {Barman},
  {Allard}, {Hauschildt}, \& {Lammer}}]{Baraffe04}
{Baraffe}, I., {Selsis}, F., {Chabrier}, G., {Barman}, T.~S., {Allard}, F.,
  {Hauschildt}, P.~H., \& {Lammer}, H. 2004, \aap, 419, L13

\bibitem[{{Ben-Jaffel}(2007)}]{Ben-Jaffel07}
{Ben-Jaffel}, L. 2007, \apjl, 671, L61

\bibitem[{{Cecchi-Pestellini} {et~al.}(2009){Cecchi-Pestellini}, {Ciaravella},
  {Micela}, \& {Penz}}]{Cecchi-Pestellini09}
{Cecchi-Pestellini}, C., {Ciaravella}, A., {Micela}, G., \& {Penz}, T. 2009,
  \aap, 496, 863

\bibitem[{{Cohen} {et~al.}(2010){Cohen}, {Drake}, {Kashyap}, {Sokolov}, \&
  {Gombosi}}]{Cohen10}
{Cohen}, O., {Drake}, J.~J., {Kashyap}, V.~L., {Sokolov}, I.~V., \& {Gombosi},
  T.~I. 2010, \apjl, 723, L64

\bibitem[{{Garcia Mu{\~n}oz}(2007)}]{Garcia-Munoz07}
{Garcia Mu{\~n}oz}, A. 2007, Plan. and Sp. Sci., 55, 1426

\bibitem[{{Glocer} {et~al.}(2007){Glocer}, {Gombosi}, {T{\'o}th}, {Hansen},
  {Ridley}, \& {Nagy}}]{Glocer07}
{Glocer}, A., {Gombosi}, T.~I., {T{\'o}th}, G., {Hansen}, K.~C., {Ridley},
  A.~J., \& {Nagy}, A. 2007, \jgr, 112, A01304

\bibitem[{{Glocer} {et~al.}(2012){Glocer}, {Kitamura}, {Toth}, \&
  {Gombosi}}]{Glocer12}
{Glocer}, A., {Kitamura}, N., {Toth}, G., \& {Gombosi}, T. 2012, Journal of
  Geophysical Research (Space Physics), 117, 4318

\bibitem[{{Glocer} {et~al.}(2009){Glocer}, {T{\'o}th}, {Gombosi}, \&
  {Welling}}]{Glocer09}
{Glocer}, A., {T{\'o}th}, G., {Gombosi}, T., \& {Welling}, D. 2009, \jgr, 114,
  5216

\bibitem[{{Gombosi}(2004)}]{Gombosi04}
{Gombosi}, T.~I. 2004, {Physics of the Space Environment} ({Cambridge
  University Press, Cambridge, UK})

\bibitem[{{Hubbard} {et~al.}(2007){Hubbard}, {Hattori}, {Burrows}, {Hubeny}, \&
  {Sudarsky}}]{Hubbard07}
{Hubbard}, W.~B., {Hattori}, M.~F., {Burrows}, A., {Hubeny}, I., \& {Sudarsky},
  D. 2007, Icarus, 187, 358

\bibitem[{{Khazanov} {et~al.}(1997){Khazanov}, {Liemohn}, \&
  {Moore}}]{khazanov97}
{Khazanov}, G.~V., {Liemohn}, M.~W., \& {Moore}, T.~E. 1997, \jgr, 102, 7509

\bibitem[{{Lammer} {et~al.}(2003){Lammer}, {Selsis}, {Ribas}, {Guinan},
  {Bauer}, \& {Weiss}}]{Lammer03}
{Lammer}, H., {Selsis}, F., {Ribas}, I., {Guinan}, E.~F., {Bauer}, S.~J., \&
  {Weiss}, W.~W. 2003, \apjl, 598, L121

\bibitem[{{Lecavelier Des Etangs} {et~al.}(2010){Lecavelier Des Etangs},
  {Ehrenreich}, {Vidal-Madjar}, {Ballester}, {D{\'e}sert}, {Ferlet},
  {H{\'e}brard}, {Sing}, {Tchakoumegni}, \& {Udry}}]{LecavelierDesEtangs10}
{Lecavelier Des Etangs}, A., {Ehrenreich}, D., {Vidal-Madjar}, A., {Ballester},
  G.~E., {D{\'e}sert}, J.-M., {Ferlet}, R., {H{\'e}brard}, G., {Sing}, D.~K.,
  {Tchakoumegni}, K.-O., \& {Udry}, S. 2010, \aap, 514, A72

\bibitem[{{Lemaire}(1972)}]{Lemaire72}
{Lemaire}, J. 1972, Journal of Atmospheric and Terrestrial Physics, 34, 1647

\bibitem[{{Lennartsson} {et~al.}(1981){Lennartsson}, {Sharp}, {Shelley},
  {Johnson}, \& {Balsiger}}]{Lennartsson81}
{Lennartsson}, W., {Sharp}, R.~D., {Shelley}, E.~G., {Johnson}, R.~G., \&
  {Balsiger}, H. 1981, \jgr, 86, 4628

\bibitem[{{Linsky} {et~al.}(2010){Linsky}, {Yang}, {France}, {Froning},
  {Green}, {Stocke}, \& {Osterman}}]{Linsky10}
{Linsky}, J.~L., {Yang}, H., {France}, K., {Froning}, C.~S., {Green}, J.~C.,
  {Stocke}, J.~T., \& {Osterman}, S.~N. 2010, \apj, 717, 1291

\bibitem[{{Mayor} {et~al.}(2003){Mayor}, {Naef}, {Pepe}, {Queloz}, {Santos}, \&
  {Udry}}]{exoplanets03}
{Mayor}, M., {Naef}, D., {Pepe}, F., {Queloz}, D., {Santos}, N., \& {Udry}, S.
  2003, {The Geneva extrasolar planet search programmes},
  {http://exoplanets.eu}

\bibitem[{{Murray-Clay} {et~al.}(2009){Murray-Clay}, {Chiang}, \&
  {Murray}}]{Murray-clay09}
{Murray-Clay}, R.~A., {Chiang}, E.~I., \& {Murray}, N. 2009, \apj, 693, 23

\bibitem[{{Nagy} {et~al.}(1986){Nagy}, {Barakat}, \& {Schunk}}]{Nagy86}
{Nagy}, A.~F., {Barakat}, A.~R., \& {Schunk}, R.~W. 1986, \jgr, 91, 351

\bibitem[{{Papaloizou} {et~al.}(2007){Papaloizou}, {Nelson}, {Kley}, {Masset},
  \& {Artymowicz}}]{Papaloizou07}
{Papaloizou}, J.~C.~B., {Nelson}, R.~P., {Kley}, W., {Masset}, F.~S., \&
  {Artymowicz}, P. 2007, Protostars and Planets V, 655

\bibitem[{{Penz} {et~al.}(2008){Penz}, {Micela}, \& {Lammer}}]{Penz08}
{Penz}, T., {Micela}, G., \& {Lammer}, H. 2008, \aap, 477, 309

\bibitem[{{Schneider}(1995)}]{exoplanet95}
{Schneider}, J. 1995, {The Extrasolar Planets Encyclopaedia},
  {http://exoplanet.eu}

\bibitem[{{Schunk} \& {Nagy}(2004)}]{SchunkNagy04}
{Schunk}, R.~W., \& {Nagy}, A.~F. 2004, {Ionospheres} ({Cambridge University
  Press, Cambridge, UK})

\bibitem[{{Showman} {et~al.}(2008){Showman}, {Menou}, \& {Cho}}]{Showman08}
{Showman}, A.~P., {Menou}, K., \& {Cho}, J.~Y.-K. 2008, in Astronomical Society
  of the Pacific Conference Series, Vol. 398, Extreme Solar Systems, ed.
  {D.~Fischer, F.~A.~Rasio, S.~E.~Thorsett, \& A.~Wolszczan}, 419

\bibitem[{{Stone} \& {Proga}(2009)}]{Stone09}
{Stone}, J.~M., \& {Proga}, D. 2009, \apj, 694, 205

\bibitem[{{Tam} {et~al.}(1998){Tam}, {Yasseen}, \& {Chang}}]{Tam98}
{Tam}, S.~W.~Y., {Yasseen}, F., \& {Chang}, T. 1998, Annales Geophysicae, 16,
  948

\bibitem[{{Tam} {et~al.}(1995){Tam}, {Yasseen}, {Chang}, \& {Ganguli}}]{Tam95}
{Tam}, S.~W.~Y., {Yasseen}, F., {Chang}, T., \& {Ganguli}, S.~B. 1995, \grl,
  22, 2107

\bibitem[{{Tian} {et~al.}(2005){Tian}, {Toon}, {Pavlov}, \& {De
  Sterck}}]{Tian05}
{Tian}, F., {Toon}, O.~B., {Pavlov}, A.~A., \& {De Sterck}, H. 2005, \apj, 621,
  1049

\bibitem[{{Trammell} {et~al.}(2011){Trammell}, {Arras}, \& {Li}}]{Trammell11}
{Trammell}, G.~B., {Arras}, P., \& {Li}, Z.-Y. 2011, \apj, 728, 152

\bibitem[{{Vidal-Madjar} {et~al.}(2003){Vidal-Madjar}, {Lecavelier des Etangs},
  {D{\'e}sert}, {Ballester}, {Ferlet}, {H{\'e}brard}, \&
  {Mayor}}]{Vidal-Madjar03}
{Vidal-Madjar}, A., {Lecavelier des Etangs}, A., {D{\'e}sert}, J.-M.,
  {Ballester}, G.~E., {Ferlet}, R., {H{\'e}brard}, G., \& {Mayor}, M. 2003,
  \nat, 422, 143

\bibitem[{{Vidal-Madjar} {et~al.}(2008){Vidal-Madjar}, {Lecavelier des Etangs},
  {D{\'e}sert}, {Ballester}, {Ferlet}, {H{\'e}brard}, \&
  {Mayor}}]{Vidal-Madjar08}
---. 2008, \apjl, 676, L57

\bibitem[{{Yelle}(2004)}]{Yelle04}
{Yelle}, R.~V. 2004, Icarus, 170, 167

\end{thebibliography}

%%%%%%%%%%%%%%%%%%%%%%%%%%%%%%%%%%%%%%%%%%%%%%%%%%%%%%%%%%%%%%%%%%%%%%%%%%%%%%%
% Tables
%%%%%%%%%%%%%%%%%%%%%%%%%%%%%%%%%%%%%%%%%%%%%%%%%%%%%%%%%%%%%%%%%%%%%%%%%%%%%%%

%%%%%%%%%%%%%%%%%%%%%%%%%%%%%%%%%%%%%%%%%%%%%%%%%%%%%%%%%%%%%%%%%%%%%%%%%%%%%%
% Figures
%%%%%%%%%%%%%%%%%%%%%%%%%%%%%%%%%%%%%%%%%%%%%%%%%%%%%%%%%%%%%%%%%%%%%%%%%%%%%%

\clearpage

\clearpage

\clearpage

\end{document}